# Temperature dependence of the solid-liquid interface free energy of Ni and Al from molecular dynamics simulation of nucleation

Yang Sun[1], Feng Zhang[1*], Huajing Song[1],
Mikhail I. Mendelev[1*], Cai-Zhuang Wang[1,2], Kai-Ming Ho[1,2,3]

[1]Ames Laboratory, US Department of Energy, Ames, Iowa 50011, USA

[2]Department of Physics, Iowa State University, Ames, Iowa 50011, USA

[3]Hefei National Laboratory for Physical Sciences at the Microscale and Department of Physics, University of Science and Technology of China, Hefei, Anhui 230026, China

**ABSTRACT**

The temperature dependence of the solid-liquid interfacial free energy, $\gamma$, is investigated for Al and Ni at the undercooled temperature regime based on a recently developed persistent-embryo method. The atomistic description of the nucleus shape is obtained from molecular dynamics simulations. The computed $\gamma$ shows a linear dependence on the temperature. The values of $\gamma$ extrapolated to the melting temperature agree well with previous data obtained by the capillary fluctuation method. Using the temperature dependence of $\gamma$, we estimate the nucleation free energy barrier in a wide temperature range from the classical nucleation theory. The obtained data agree very well with the results from the brute-force molecular dynamics simulations.

## I. INTRODUCTION

The solid-liquid interfacial (SLI) free energy, $\gamma$, plays a fundamental role in crystal nucleation and growth process[1]. It is also a key parameter required to model the formation of solidification microstructures[2]. Despite its importance, the measurement of the SLI free energy is extremely difficult in experiments. Therefore, computer simulation, which provides detailed atomistic information, remains heavily employed to quantitatively investigate $\gamma$.

A well-established method to compute $\gamma$ is the capillary fluctuation method (CFM)[3] which measures the SLI stiffness based on capillary wave theory [4,5]. While CFM makes an accurate determination of $\gamma$, it is only available at the melting point $T_m$ and usually computationally expensive[6]. To obtain $\gamma$ at other temperatures, Laird and co-workers further extend the CFM results along the pressure-temperature coexistence curve using the "Gibbs-Cahn integration" method [7]. However, the temperature dependence of $\gamma$ at $p=0$ remains unclear. Moreover, in the case when several crystal phases compete with each other, a large pressure can trigger a nucleation of the phase which was metastable at $p=0$. On the other hand, one can make an indirect measurement of the SLI free energy from nucleation simulation with the classical nucleation theory (CNT)[8,9]. This method utilizes the results of molecular dynamics (MD) simulations where the critical nucleus was actually observed. While the method is in principle reliable (see details below), the accuracy strongly depends on the measurement of the size and shape of the critical nucleus[10]. In particular, this method faces the well-known difficulty associated with the fact that the nucleation is usually too rare event. Recently we developed a persistent-embryo method (PEM)[11] to overcome this problem in moderately undercooled liquids. With the PEM, one can observe the actual fluctuations of the large critical nucleus without any biasing. In this work, using the PEM, we determined the average nucleus shape for two fcc crystals, Al and Ni, in the moderately undercooled regime. Then the temperature dependence of the SLI free energy was obtained in the framework of the CNT. These data were used in turn to predict the free energy barrier in a wide temperature range for both systems.

---

[*]Email: fzhang@ameslab.gov (F.Z.)
[*]Email: mendelev@ameslab.gov (M.I.M.)

The rest of the paper is organized as follows: in Section II, we will introduce the persistent embryo method and provide the simulation details. In Section III, we will present the obtained temperature dependences of SLI free energy for Al and Ni. In Section IV, we will show the obtained SLI free energy data lead to the nucleation barriers in agreement with the data determined using a very different technique. In Section V, we will discuss the obtained results and we will provide the summary in Section VI.

## II. PERSISTENT EMBRYO METHOD

According to the CNT [1], a homogeneous nucleation involves a formation of the critical nucleus in the undercooled liquid. The formation of such a nucleus is governed by two factors. The first one is the thermodynamic driving force towards the lower-free-energy bulk crystal. This term is negative and proportional to the number of atoms in the nucleus. The other is the energy penalty for creating an interface between the nucleus and the liquid. This term is positive and proportional to the area of the interface. Therefore, the excess free energy to form a nucleus with $N$ atoms is

$$\Delta G = N\Delta\mu + A\gamma, \quad (1)$$

where $\Delta\mu$ ($< 0$) is the chemical potential difference between the bulk solid and liquid, $\gamma$ is the solid-liquid interfacial free energy, and $A$ is the interface area which can be evaluated as $A = s(N/\rho_c)^{2/3}$, where $\rho_c$ is the crystal density and $s$ is a shape factor. The competition between the bulk and interface terms leads to a nucleation barrier $\Delta G^*$ when the nucleus reaches the critical size $N^*$, i.e. $\frac{\partial \Delta G(N^*)}{\partial N} = 0$, and

$$\Delta G^* = \frac{4s^3\gamma^3}{27|\Delta\mu|^2\rho_c^2}. \quad (2)$$

The CNT assumes the spherical shape ($s_{CNT} \equiv \sqrt[3]{36\pi}$) for the nucleus to relate $\Delta G^*$ with $\gamma$ and $\Delta\mu$. This assumption can be lifted by introducing the shape factor $s$, assuming that the averaged shape of the sub-critical nucleus does not change at the critical size. Mathematically, the interfacial free energy density $\gamma$ and the shape factor $s$ in Eq. (2), which are both difficult to compute, can be replaced by the critical nucleus size $N^*$ at the critical point[11] based on the relation

$$\gamma = \frac{3}{2s}|\Delta\mu|\rho_c^{2/3}N^{*\frac{1}{3}}, \quad (3)$$

resulting in $\Delta G^* = \frac{1}{2}|\Delta\mu|N^*$. According to Eq. (3), four quantities ($\rho_c$, $\Delta\mu$, $N^*$, and $s$) are needed to obtain from the MD to calculate the interfacial free energy $\gamma$ at a given temperature. The determination of the crystal density, $\rho_c$, is trivial. The chemical potential difference, $\Delta\mu$, can be calculated by integrating the Gibbs-Helmholtz equation from the undercooling temperature to the melting point[12]. The determination of the critical nucleus size $N^*$ and the shape factor can be obtained from the PEM simulations which will be described in detail below.

The PEM utilizes the main CNT concept that homogeneous nucleation happens via the formation of the critical nucleus in the undercooled liquid. The PEM allows efficient sampling of the nucleation process by preventing a small crystal embryo (with $N_0$ atoms which is much smaller than the critical nucleus) from melting using external spring forces[11]. This removes long periods of ineffective simulation where the system is very far away from forming a critical nucleus. As the embryo grows, the harmonic potential is gradually weakened and is completely removed when the cluster size reaches a sub-critical threshold $N_{sc}$ ($< N^*$). During the simulation, the harmonic potential only applies to the original $N_0$($< N_{sc}$) embryo atoms. The spring constant of the harmonic potential decreases with increasing the nucleus size as $k(N) = k_0 \frac{N_{sc}-N}{N_{sc}}$ if $N < N_{sc}$ and $k(N) = 0$, otherwise. This strategy ensures the system is unbiased at the critical point such that a reliable critical nucleus is obtained. If the nucleus melts below $N_{sc}$ ($< N^*$) the harmonic potential is gradually enforced preventing the complete melting of the embryo. When the nucleus reaches the critical size, it has equal chance to melt or to further grow causing fluctuations about $N^*$. As a result, the $N(t)$ curve tends to display a plateau during the critical fluctuations, giving a unique signal to detect the appearance of the critical nucleus. In addition, multiple plateaus can be collected before a critical nucleus eventually grows, allowing sufficient statistical analysis of nuclei's size and shape.

All MD simulations in the present study were performed using the GPU-accelerated LAMMPS code[13–15]. The interatomic interaction was modelled using the Finnis-Sinclair potentials [16] developed for the Ni[17] and Al[12]. During the MD simulation, the NPT ensemble was applied with Nose-Hoover thermostats. The damping time in the Nose-Hoover thermostat is set as $\tau = 0.1\ ps$ which is frequent enough for the heat dissipation during the crystallization (see the Supplementary Material). The time step of the simulation was $1.0\ fs$. The simulation cell contained up to 32,000 atoms which is at least 20 times larger than the critical nucleus size. This setting ensures the effect of pressure change during the nucleation is minimal to the entire simulation box (see the Supplementary Material).

To identify the nucleus size during the MD simulation, we used the bond-orientational order (BOO) parameter[18,19]. In this approach, one first defines the correlation between the structures of two neighbor atoms $i$ and $j$ as

$$S_{ij} = \sum_{m=-6}^{6} q_{6m}(i) \cdot q_{6m}^*(j) ,\ (4)$$

where

$$q_{6m}(i) = \frac{1}{N_b(i)} \sum_{j=1}^{N_b(i)} Y_{lm}(\vec{r}_{ij})\ (5)$$

is the Steinhardt parameter, $Y_{lm}(\vec{r}_{ij})$ are the spherical harmonics, $N_b(i)$ is the number of nearest neighbors of atom $i$ and $\vec{r}_{ij}$ is the vector connecting it with its neighbor $j$. Two neighboring atoms $i$ and $j$ are considered to be connected when $S_{ij}$ exceeds a threshold $S_c$. To choose a reasonable value of $S_c$, Espinosa *et al.*'s suggested an "equal mislabeling" method[20] by plotting the population of mislabeled atoms in the bulk solid and liquid as a function of the threshold values. As shown in Fig. 1(a), the crossing point of the mislabeling curves of the bulk liquid and solid phases is chosen as the threshold, $S_c$, to provide that the probability of mislabeling atoms in the bulk liquid as solid-like atoms is the same as the probability of mislabeling atoms in the bulk solid as liquid-like atoms.

This approach works very well when one needs to detect "solid" atoms within a bulk liquid. However, it tends to mislabel "solid" atoms at the cluster interface. To account for that, one can determine how many solid-like neighbors an atom has. Figure 1(b) shows that this quantity, $\xi$, is quite different for majority of atoms in the bulk solid and liquid phases and the number of mislabeled atoms is very small (see the insert in this figure). Intuitively, it is natural to choose the threshold value, $\xi_c$, to be 6 for FCC-liquid interfaces. This approach is quite sufficient for the PEM which requires on-the-fly identification of solid-like atoms during the MD simulation. However, recent study shows that the choice of $\xi_c$ considerably affects the value of $N^*$ determined from the MD snapshots[21]. We will return to this issue in Section V.

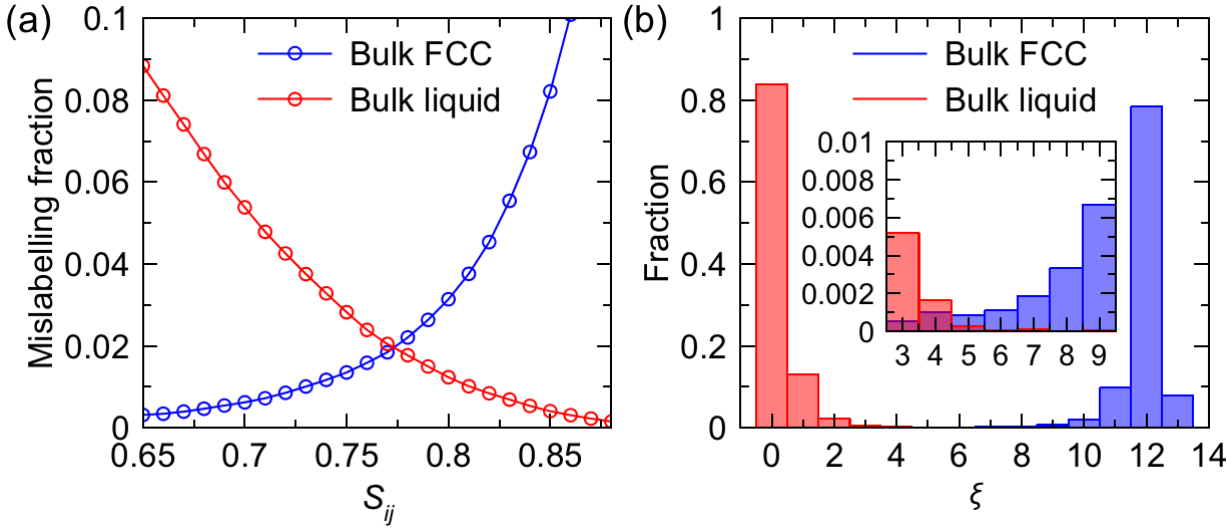

Fig. 1. Determination of the threshold to distinguish solid-like and liquid-like atoms. (a) Population of mislabeled atoms by different threshold values in bulk Ni crystal and liquid at 1430 K. (b) Population of connections number per atom in bulk Ni crystal and liquid at 1430K. The insert zooms in the region of $\xi$ from 3 to 9.

## III. TEMPERATURE DEPENDENCE OF THE SLI FREE ENERGY

Figure 2(a) shows a typical PEM simulation. The plateau indicates the appearance of the critical nucleus. Therefore the critical size $N^*$ can be directly measured by averaging the size at the plateau[11]. To make a statistically sound description of the nucleus shape, we first averaged the nucleus by superposing the configurations collected in a short time interval ($\Delta t_0 = 10\ ps$) during the plateau. As shown in Fig. 2(b), the superposed configuration shows a clear non-spherical nucleus shape. Since the crystalline order fades at the interfacial region, it results in a less dense atomic distribution at the outer shell of the nucleus. In order to see the averaged nucleus shape more clearly, a Gaussian smearing scheme [22–24] was applied to convert the atomic distribution into the atomic density in the 3D space. By applying a fast-clustering algorithm[25] on the density profile, we were able to extract the high-density points, which are essentially the as-formed crystalline sites. Then the crystalline sites, which were occupied in at-least half of the snapshots collected during the time interval $\Delta t_0$, were used to construct the surface of the nucleus by the geometric surface reconstruction method[26] integrated in the OVITO software package[27] as shown in Fig. 2(b). Finally, the shape factor, $s$ was computed based on the surface area $A$ and the volume $V$ of the polyhedron computed from OVITO as $s = A/V^{2/3}$. Figure 2(c) shows the measured shape factor and the critical nucleus size as functions of temperature for Ni. The shape factor clearly demonstrates a non-spherical shape. However, while the critical nucleus size dramatically increases with the increase of the temperature, the shape factor shows only a slight decrease.

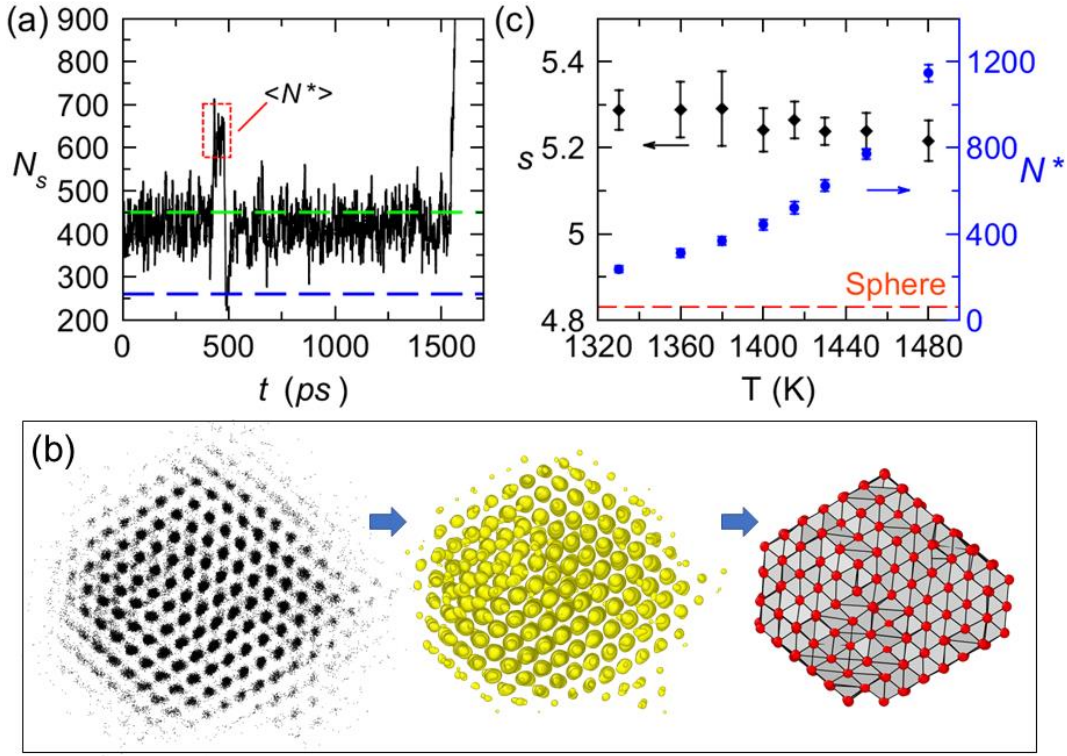

Fig. 2 (a) Nucleus size as a function of time in a typical PEM simulation for Ni at 1430 K. Blue dashed line shows the size of the embryo, $N_0$, and the green dashed line shows the threshold $N_{sc}$ to remove the spring on the embryo atoms. The box (red) indicates the plateaus of the critical nucleus. (b) From left to right: Superposed fcc nucleus configurations obtained from the plateau in the PEM simulation; the density contour plot corresponding to the atomic distribution in the superposed configuration; the surface of the polyhedron constructed by the high-density points. (c) The measured shape factors (black) and the size (blue) of the critical nucleus for Ni as a function of the temperature. The error bars are obtained by measuring the shape factors of different critical nucleus collected from PEM simulations. The dash line indicates the shape factor under spherical shape assumption.

With the measured shape factor and the critical size, the interfacial free energy $\gamma$ can be calculated by Eqn. (3). Figure 3 shows the obtained data for both Ni and Al. In both systems, the interfacial free energy $\gamma$ shows a nearly linear dependence on the temperature. Therefore, we fit the data with a linear relation to the temperature and extrapolate to the melting point. Figure 3 shows that within the accuracy of the measurement, the extrapolated interfacial free energies agree very well with the data obtained by CFM[3] for both Al and Ni[28].

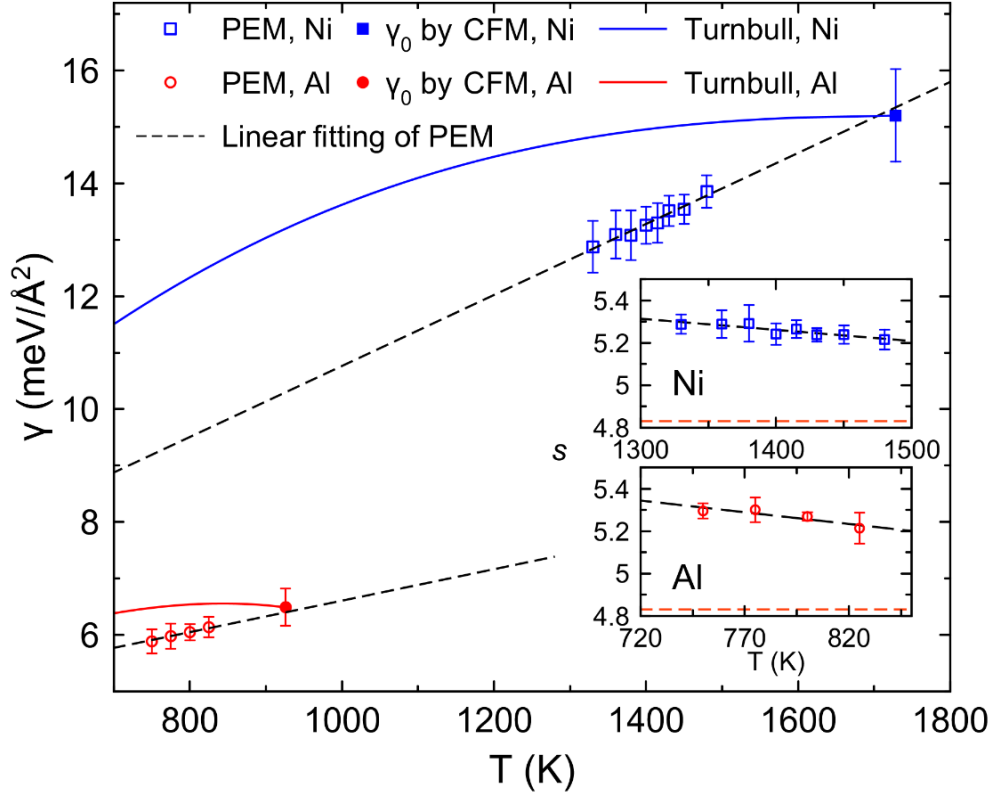

Fig. 3 The interfacial free energy as a function of the temperature for Ni and Al. The open squares and circles are the data obtained using the PEM. The error bar of PEM results are obtained by the error propagation of Eqn. (3) as $\sigma_\gamma = \frac{3}{2s}|\Delta\mu|\rho_c^{2/3} N^{*\frac{1}{3}}\sqrt{\frac{\sigma_s^2}{s^2} + \frac{1}{9}\frac{\sigma_{N^*}^2}{N^{*2}}}$, where $\sigma_s$, $\sigma_{N^*}$ are the statistic uncertainties of the measurement of $s$ and $N^*$ in the PEM simulations. The filled square and circle are the data obtained at the melting points of Ni and Al using the CFM [28]. The dash lines are the linear fitting and extrapolations of the PEM data ($\gamma_{\text{Ni}} = 4.475 + 0.006290T$ ($meV/\text{Å}^2$) and $\gamma_{\text{Al}} = 3.819 + 0.002788T$ ($meV/\text{Å}^2$)). The solid lines are obtained from the Turnbull correlation. The insert shows linear fitting of the measured shape factor as a function of the temperature for both systems. The red dashed in the insert shows the shape factor of spherical assumption as the reference.

**IV. CALCULATION OF THE NUCLEATION BARRIER**

A straightforward application of the temperature dependence of the interfacial free energy $\gamma$ is to estimate the free energy barrier at very small and very large supercoolings where the PEM cannot be applied. The case of very small supercooling is interesting because it corresponds to the experimental conditions of solidification. The only way to judge about the reliability of the calculations here is to compare with the experimental data although both experimental and computational data will be affected by the factors not related to the CNT (e.g., the quality of the employed semi-empirical potential in the case of simulation or the presence of impurities in the case of experiment). The case of very large supercooling in the case of pure metals is interesting because the nucleation rate can be directly obtained from the MD simulation. In this case, the quality of the employed semi-empirical potential is not an issue. However, the extrapolation to this temperature range may not work because of several other issues. For example, the temperature dependence of the SLI free energy can be different than the one observed at higher temperatures. Another issue is associated with the fact that the critical nucleus at low temperatures becomes so small that the entire CNT concept may not be applicable.

In the extrapolation of the nucleation barriers (see Eqn. 2), we used a linear fitting for the temperature dependences of the SLI free energy and the shape factor (see Fig. 2). The obtained temperature dependences of the nucleation barriers are shown in Fig. 4. The obtained temperature dependences well describe our PEM data, which was expected because these dependences were obtained by fitting to the PEM data. The question is if these dependences can be useful to predict the nucleation barrier in a temperature range where the PEM is not applicable. In the case of Ni, the nucleation barrier for the same semi-empirical potential was obtained at $T$=1180 K [29] using the combination of the mean first-passenger time (MFPT) method [30–32] and the Fokker-Planck equation [30,31,33] directly from an unbiased MD simulation[34,35]. In the present work, we used exactly the same approach to obtain the nucleation barrier for Al at $T$=580 K. Figure 4 shows that the obtained MFPT data are in excellent agreement with the data we obtained using the temperature dependences of the SLI free energies.

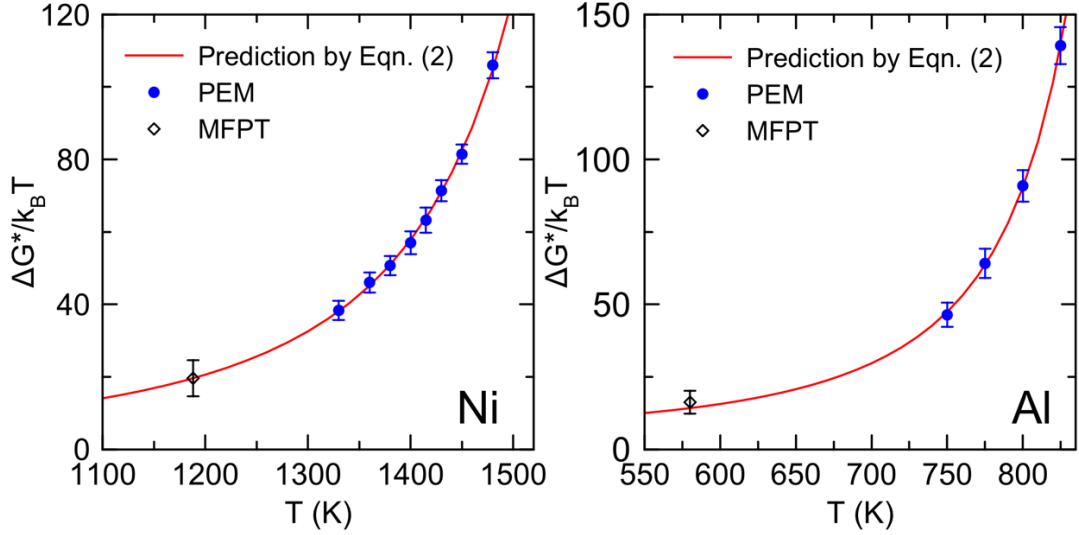

Fig. 4 The predicted temperature dependence of the nucleation barrier for Ni and Al. The PEM data of Ni is from Ref. 11 and the MFPT data of Ni is from Ref. [29]. The PEM and MFPT data of Al is measured in the current work. The error bars are obtained as $\sigma_{\Delta G^*} = \frac{1}{2}|\Delta \mu|\sigma_{N^*}$, where $\sigma_{N^*}$ are the uncertainties of the measurement of $N^*$ in the PEM simulations.

## V. DISCUSSION

In the present study we obtained the temperature dependence of the SLI free energy at the moderate undercooling range where other existing techniques are not applicable. Therefore, to validate the obtained results we extrapolated the obtained temperature dependences to the temperatures where well-established methods can be applied. Figure 3 shows that the extrapolation to the melting temperature very well agrees with the CFM data. It should be noted that contrary to the CFM which provides the SLI free energy as function of the interface orientation, in the present study we obtained the SLI free energy averaged over all orientation using the CNT framework (see Eqn. 3). Therefore, we compare the current results to the $\gamma_0$ value from the CFM (see Eqn. 1 in Ref. 3). This was reasonable for pure Ni and Al since the anisotropy of the SLI free energy is not very large for the pure fcc metals[36,37] at least at the melting temperature. Moreover, the PEM provides ample statistics to measure the shape of the nucleus in the temperature range where it is applicable and in the present work, we did not observe very large deviation from the spherical nucleus shape.

However, one should be cautious in the interpretation of the SLI free energy value obtained from the PEM in the case crystal phase with very anisotropic SLI free energy (e.g., see Fig. 10 in Ref. [38]).

Another possibility to validate our results was to extrapolate the obtained temperature dependences to low temperatures and compare the obtained nucleation barrier free energies with the data obtained from the brute-force MD simulations. The obtained excellent agreement is rather surprising because it suggests that the CNT still works at these temperatures in spite of the fact that critical nucleus size (only tens of atoms[29]) is so small that it is not really possible to distinguish between the bulk and the interface regions within a nucleus. In this case, even the concept of the SLI free energy is not clear. Yet, one can always describe the change in the free energy associated with the nucleus formation as the sum of two contributions: the product of the difference in the bulk free energy per atom and the number of atoms in the nucleus and a contribution, which accounts for the nucleus interface. The latter can be treated as the flat interface free energy corrected for the high interface curvature (e.g., see Ref. [35,39,40]). In fact, this is the quantity we obtained from the PEM. At high temperatures, where the nucleus is large and the correction for the high interface curvature is negligible we obtained a good agreement with the flat interface free energy data from the CFM. At low temperatures, we obtained a good agreement with the brute-force MD simulation data but the quantity we extracted includes not just the flat SLI free energy but also corrections associated with the SLI curvature. The authors of Ref. 41 argued that namely these corrections explain why the value of the SLI free energy obtained from the seeding simulations is always below that estimated from the Turnbull correlation [42] which was proposed in Ref. 34 to use to estimate the temperature dependence of the SLI free energy. The temperature dependences of the SLI free energy obtained in the present study are also below the predictions based on the Turnbull correlation (see Fig. 3).

The main source of the uncertainty in the determined value of the SLI free energy comes from the uncertainty in determination of the number of atoms in the critical crystal cluster, $N^*$. This quantity can be rather sensitive to the choice of order parameters as has been noted in [21,43] and can be seen in Fig. 5. In addition to the BOO parameter we also employed the cluster-alignment (CA) method[23] in which minimal root-mean-square deviations (RMSD) between the atom cluster and the perfect packing templates such as FCC, HCP and BCC polyhedral are calculated for crystal-structure recognition. Interestingly that the CA order parameter leads to almost identical results comparing to the use of the BOO parameter with $\xi_c = 6$ which was assumed to be the most reasonable value.

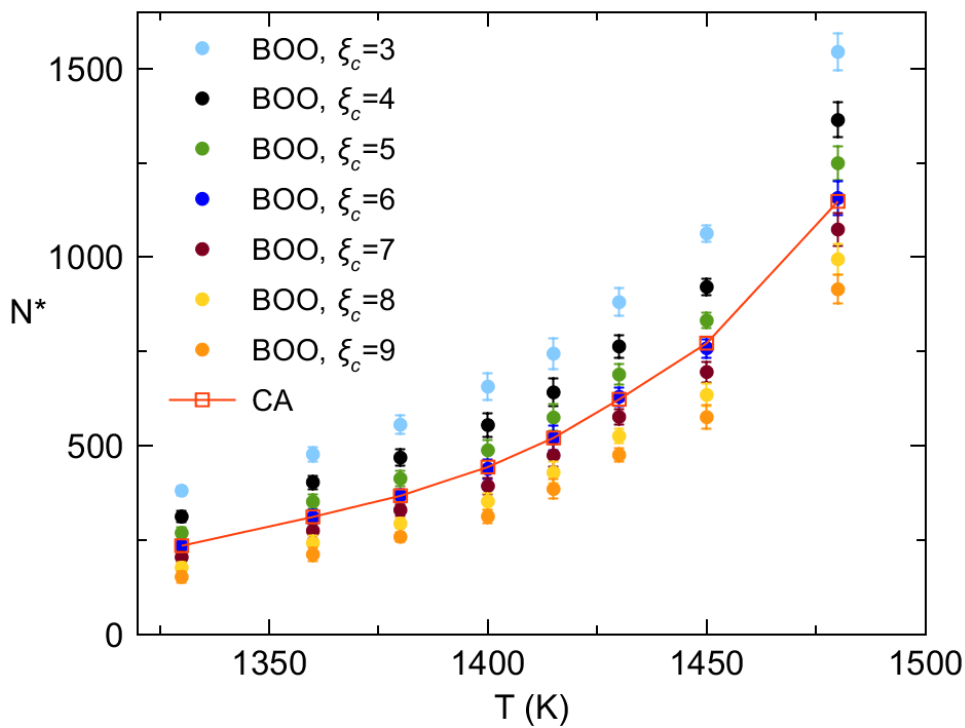

Fig. 5. Dependence of the critical nucleus size in Ni determined from MD simulation on the choice of the order parameter (BOO or CA) or the threshold value in the BOO parameter.

Figure 6 shows how the uncertainty in $N^*$ caused by the choice of the order parameters propagates in the uncertainty of the SLI free energy determined within the present study. A vivid systematic difference can be seen. However, it is important that the temperature dependence remains qualitatively the same: no matter what order parameter we used the obtained temperature dependence was linear. What is even more important is that all lines come to almost the same point which is in excellent agreement with the CFM value of the SLI free energy.

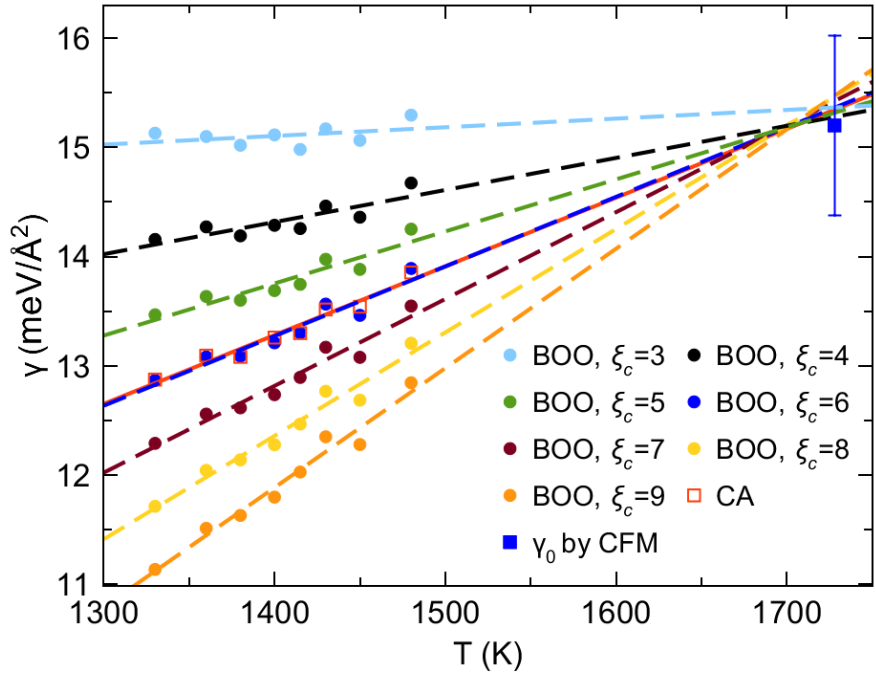

Fig. 6. The temperature dependence of the SLI free energy in Ni calculated with the critical nucleus sizes determined using different order parameters. The dash lines indicate the linear fitting of the dots/square with the same color.

The extrapolation to low temperatures is shown in Fig. 7. The obtained results indeed depend on the choice of the order parameter and extreme choices can lead to considerable overestimations or underestimations of the nucleation barrier. However, the reasonable choice of the threshold value in the BOO parameter ($\xi_c = 6$) or using the CA order parameter provide an excellent agreement with the brute-force MD simulation. Moreover, using slightly different values of the threshold value in the BOO parameter ($\xi_c = 5$ or $\xi_c = 7$) lead to the variations in the nucleation barrier value obtained by extrapolation of the PEM data within uncertainty of the brute-force MD simulation data.

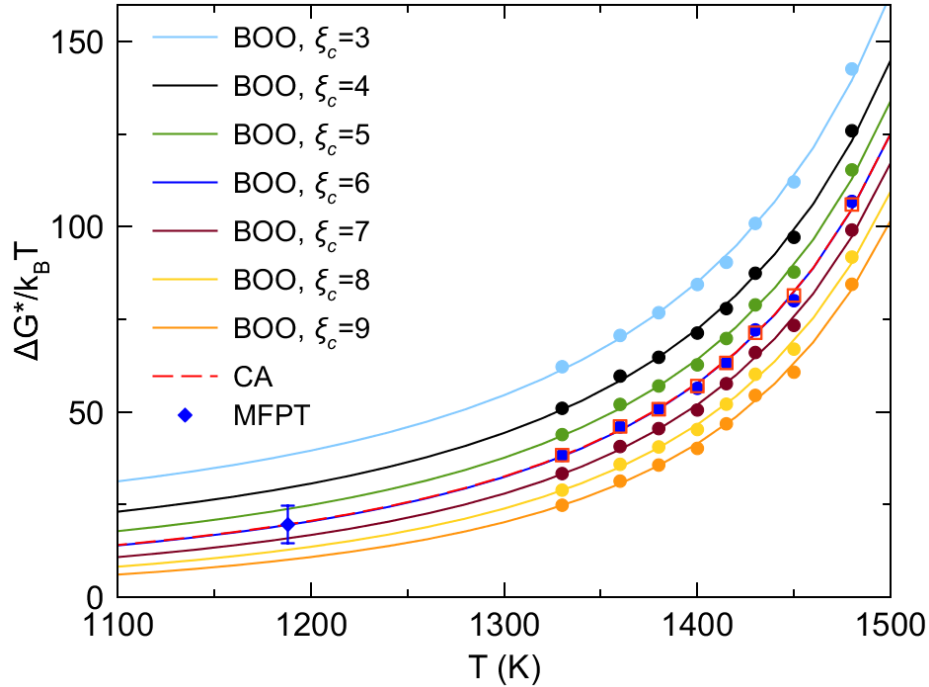

Fig. 7. The temperature dependence of the nucleation barrier in Ni calculated with the critical nucleus sizes determined using different order parameters.

**VI. CONCLUSIONS**

In summary, based on the atomic configurations of critical nucleus obtained by the PEM method, we estimate the shape factor of the nucleus and find both critical nucleus of Ni and Al deviate from the spherical shape. With the framework of CNT, we obtain a nearly linear temperature dependence of the orientation-averaged interfacial free energy for both Ni and Al. Using this temperature dependence, we predict the free energy barrier in a wide temperature range, which shows a good agreement with the value obtained from brute-force MD simulations.

**SUPPLEMENTARY MATERIAL**

See supplementary material for the latent heat dissipation, the analysis of the statistical uncertainty and the pressure on the nucleus.

**Acknowledgements**

Work at Ames Laboratory was supported by the US Department of Energy, Basic Energy Sciences, Materials Science and Engineering Division, under Contract No. DE-AC02-07CH11358, including a grant of computer time at the National Energy Research Supercomputing Center (NERSC) in Berkeley, CA. K.M.H. acknowledges support from USTC Qian-Ren B (1000-Talents Program B) fund. The Laboratory Directed Research and Development (LDRD) program of Ames Laboratory supported the use of GPU computing.

# Supplementary Material for "Temperature dependence of the solid-liquid interface free energy of Ni and Al from molecular dynamics simulation of nucleation"

Yang Sun[1], Feng Zhang[1*], Huajing Song[1],
Mikhail I. Mendelev[2*], Cai-Zhuang Wang[1,2], Kai-Ming Ho[1,2,3]

[1]Ames Laboratory, US Department of Energy, Ames, Iowa 50011, USA

[2]Department of Physics, Iowa State University, Ames, Iowa 50011, USA

[3]Hefei National Laboratory for Physical Sciences at the Microscale and Department of Physics, University of Science and Technology of China, Hefei, Anhui 230026, China

## I. THE LATENT HEAT DISSIPATION

During the solidification the latent heat is generated and the interface temperature can be different from the thermostat set point. This effect can be very important in the case of moving solid-liquid interface (SLI) as was shown in Ref. 1. However, this effect should be negligible in the present PEM simulations where the critical nucleus size was determined from the MD simulation. In this case, the interface does not move during the time of the plateaus from which we used to determine $N^*$. Therefore, no heat generation is anticipated.

To further check whether there is a temperature difference at the solid-liquid interface, we determined the local temperature in the nucleus and its surrounding area for Ni at 1430 K. To compute the local temperature, we first average the kinetic energy $\bar{E}_k$ over the whole critical plateaus (every 10 $fs$ for 40 $ps$ in total) for each atom. Then the local temperature within the shell with the distance $R$ to the center of the nucleus was computed by averaging $\bar{E}_k$ of the atoms in the shell from $R$ to $R + \Delta R$ as $\bar{T} = \frac{2\langle \bar{E}_k \rangle}{3k_B}$. The obtained local temperature as a function of $R$ is shown in Fig. S1. We repeated the calculations for another independent sample. Note that the averaged radius of the critical nucleus for Ni at 1430 K is 12 Å. The examination of Fig. S1 does not reveal any special features at the critical nucleus interface. Therefore, we conclude that the latent heat generated during the nucleation had sufficient time to dissipate.

---

*Email: fzhang@ameslab.gov (F.Z.)
*Email: mendelev@ameslab.gov (M.I.M.)

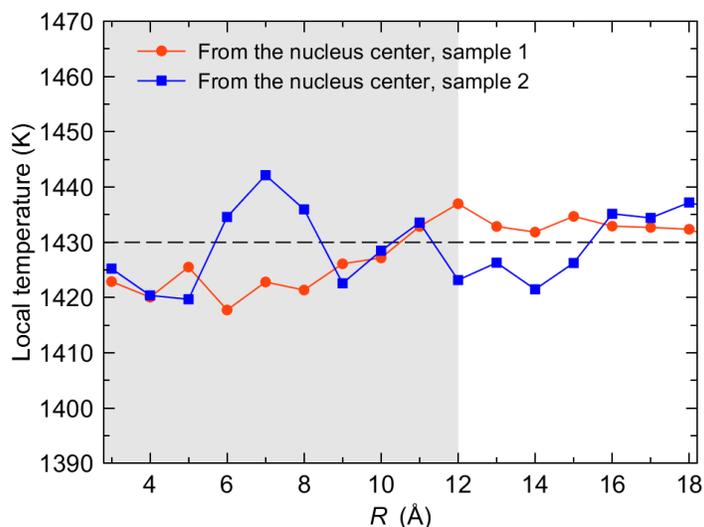

Fig. S1 The local temperature as a function of the radius to the nucleus center of mass for Ni at 1430K. The dashed line shows the target temperature of the simulation. The grey shadow indicates the average radius of the nucleus.

Next, we exam whether the Nose-Hoover relaxation time is longer than the timescale over which the size of the nucleus fluctuates. We note that the thermostat is applied to the entire simulation cell rather than to the growing nucleus region. Therefore, the thermostat can affect the obtained results only if it cannot keep up with the latent heat generated during the solidification. To test the employed thermostat we performed a simulation of the nucleus which well exceeded the critical size. In this case the heat generation is much faster than in the case of the critical nucleus and if the employed thermostat is suitable for this situation and will be even more suitable for the critical nucleus simulations. In the MD simulations reported in our paper, the damping parameter of the Nose-Hoover thermostat was set as $\tau = 0.1\ ps$ (following the recommendation of the LAMMPS developers). Figure S2(a) shows the increase of the numbers of atoms in the solid phase during the growth and the temperature in the model during this simulation. Obviously, in this case the employed thermostat is capable to keep up and the temperature does not change. Figure S2(b) shows the same simulation except $\tau = 10\ ps$. In this case, the temperature increases during the simulation, therefore, this choice of the thermostat damping parameter is not appropriate. Thus, the simulations described above justify that the choice of $\tau = 0.1\ ps$ is good enough for the heat dissipation during the crystallization. Note again, that since the nucleus size does not change significantly at the plateau, we should not expect a considerable latent heat generated during the plateau period.

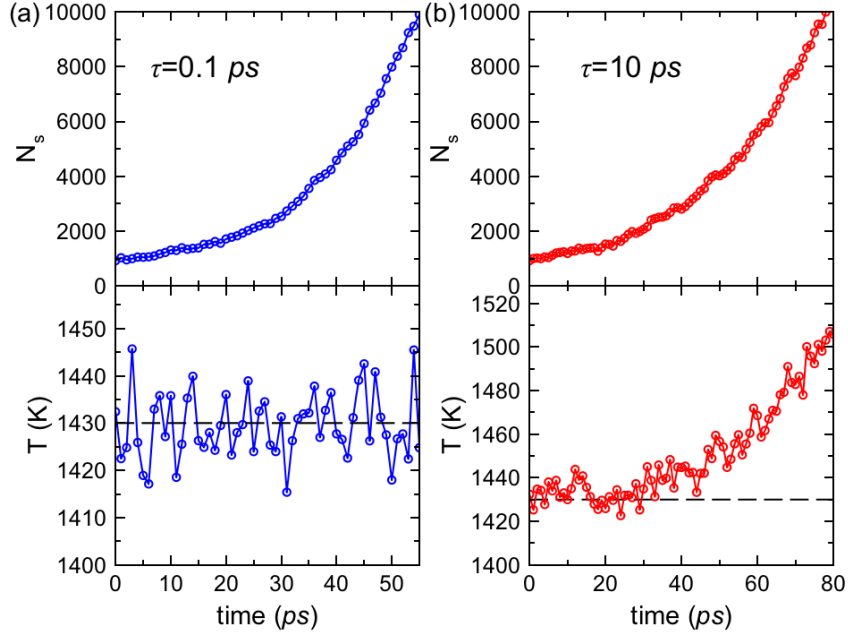

Fig. S2 (a) The growth of the supercritical nucleus of Ni at 1430K. The damping time of the Nose-Hoover thermostat is set as $\tau = 0.1\ ps$. The upper panel shows the size of the nucleus, while the lower panel monitors the temperature of the whole simulation cell. (b) The simulation starts from the same initial configuration as (a), while the damping time is set as $\tau = 10\ ps$.

## II. THE STATISTIC UNCERTAINTY

The statistic uncertainties in the current work are obtained by standard uncertainty propagations. According to Eqn. (3) in main text $\gamma = \frac{3}{2s}|\Delta\mu|\rho_c^{2/3}N^{*\frac{1}{3}}$, the statistic uncertainty of $\gamma$ (in Fig. 3 of the main text) is

$$\sigma_\gamma = \frac{3}{2s}|\Delta\mu|\rho_c^{2/3}N^{*\frac{1}{3}}\sqrt{\frac{\sigma_s^2}{s^2} + \frac{1}{9}\frac{\sigma_{N^*}^2}{N^{*2}} + \frac{\sigma_{\Delta\mu}^2}{\Delta\mu^2} + \frac{4}{9}\frac{\sigma_{\rho_c}^2}{\rho_c^2}} \quad (S1),$$

where $\sigma_s$, $\sigma_{N^*}$, $\sigma_{\Delta\mu}$, $\sigma_{\rho_c}$ are the statistic uncertainties of measuring shape factor $s$, critical nucleus size $N^*$, chemical potential difference $\Delta\mu$ and solid density $\rho_c$, respectively. According to the equation $\Delta G^* = \frac{1}{2}|\Delta\mu|N^*$, the uncertainty of the free energy barrier $\Delta G^*$ (in Fig. 4 of the main text) is

$$\sigma_{\Delta G^*} = \frac{1}{2}|\Delta\mu|N^*\sqrt{\frac{\sigma_{N^*}^2}{N^{*2}} + \frac{\sigma_{\Delta\mu}^2}{\Delta\mu^2}} \quad (S2).$$

The statistic uncertainty of the PEM simulation mainly comes from the measurement of the nucleus size and shape. The determinations of $\rho_c$ and $\Delta\mu$ from MD simulation are very accurate for pure metals. Therefore, we assume that $\sigma_{\Delta\mu} = \sigma_{\rho_c} = 0$. The uncertainties $\sigma_\gamma$ and $\sigma_{\Delta G^*}$ in the current work become

$$\sigma_\gamma = \frac{3}{2s}|\Delta\mu|\rho_c^{2/3}N^{*\frac{1}{3}}\sqrt{\frac{\sigma_s^2}{s^2} + \frac{1}{9}\frac{\sigma_{N^*}^2}{N^{*2}}}, \sigma_{\Delta G^*} = \frac{1}{2}|\Delta\mu|\sigma_{N^*}, \quad (S3)$$

which are plotted as the error bars in Fig. 3 and Fig. 4 of the main text. The systematic uncertainty, which comes from the definition of the order parameters, has been discussed in the Section V of the main text.

## III. THE PRESSURE ON THE NUCLEUS

The current simulations use a Nosé-Hoover based NPT simulation technique. Such methods ensure that pressure in the simulation box fluctuates around the target input value. In this case, a simulation box containing a crystal nucleus and a bulk fluid separated by a solid-liquid interface will have a pressure that is inhomogeneous within the system. To test whether this effect changes the pressure significantly, we examined the PEM simulation data for the plateau period shown in the Fig. 2(a) in the main text for Ni at 1430 K. To compute the local pressure in the nucleus, we defined a nucleus region by setting a box which covers most of the nucleus atom as shown in Fig. S3(b). We also set a similar box in the bulk liquid. As shown in Fig. S3(a), the local pressure in the liquid region and nucleus region did not show a significant difference when fluctuating. The averaged pressures over the time period are -0.048 GPa and 0.062 GPa for liquid region and nucleus region, which is very minimal to the entire simulation box. It is very unlike that such a small pressure can affect the obtained results but more studies are needed.

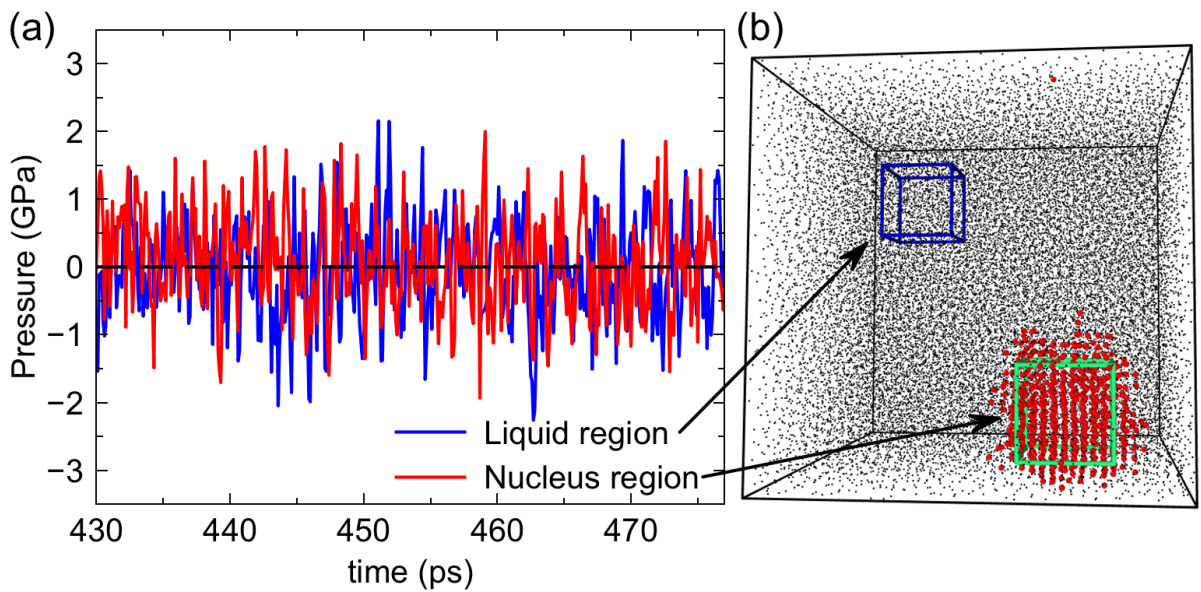

Fig. S3 (a) The pressure as a function of time during the plateau of critical nucleus for Ni at 1430K. The dashed line is a reference of $P$=0 GPa. (b) The MD snapshot at t=450 *ps*. The liquid region and nucleus region are highlighted by the blue and green boxes, respectively. The large red dots are the nucleus, while the small black dots are liquid atoms. The size of the boxes in the liquid region and nucleus region is $16 \times 16 \times 16 \text{ Å}^3$.

**Reference:**

[1] J. Monk, Y. Yang, M.I. Mendelev, M. Asta, J.J. Hoyt, and D.Y. Sun, Model. Simul. Mater. Sci. Eng. **18**, 015004 (2010).